# Machine learning identifies scale-free properties in disordered materials

Sunkyu Yu, Xianji Piao, and Namkyoo Park[*]

Photonic Systems Laboratory, Dept. of Electrical and Computer Engineering, Seoul National University, Seoul 08826, Korea

**Abstract**

The vast amount of design freedom in disordered systems expands the parameter space for signal processing, allowing for unique signal flows that are distinguished from those in regular systems. However, this large degree of freedom has hindered the deterministic design of disordered systems for target functionalities. Here, we employ a machine learning (ML) approach for predicting and designing wave-matter interactions in disordered structures, thereby identifying scale-free properties for waves. To abstract and map the features of wave behaviours and disordered structures, we develop disorder-to-localization and localization-to-disorder convolutional neural networks (CNNs). Each CNN enables the instantaneous prediction of wave localization in disordered structures and the instantaneous generation of disordered structures from given localizations. We demonstrate that CNN-generated disordered structures have scale-free properties with heavy tails and hub atoms, which exhibit an increase of multiple orders of magnitude in robustness to accidental defects, such as material or structural imperfection. Our results verify the critical role of ML network structures in determining ML-generated real-space structures, which can be used in the design of defect-immune and efficiently tunable devices.

**Introduction**

Disordered systems cover all regimes of structural phases, including periodic, quasiperiodic, and correlated or uncorrelated disordered structures, each of which has its carefully tailored strength and pattern of disorder. The classification of disorder according to microscopic structural information has thus attracted great attention in various fields, such as many-body systems[1], network science[2], and wave-matter interactions[3]. In wave physics, rich degrees of freedom in disordered systems enable exotic wave phenomena distinct from those of periodic or quasiperiodic systems, including strong[4] or weak[5] localizations, broadband responses in wave coupling[6] or absorption[7], and topological transitions with disorder-induced conductivity[8]. In particular, localization phenomena have received an extensive amount of attention as the origin of material phase transitions[9] and as the toolkit for energy confinement[3,10,11] that enables multimode lasing[12] and nanoscale sensing[13].

Traditional approaches for exploring disordered structures and their related wave behaviours have employed mapping between disordered structures and wave properties through different types of mathematical microstructural descriptors[1], such as $n$-point probability, percolation, or cluster functions. Each descriptor unveils a specific aspect of structural patterns, which enables the classification of disordered structures according to their correlations and topologies and reveals the origin of distinct wave behaviours in each class of disorder. By including the descriptors in the cost function for the optimization process, numerous inverse design methods have also been developed for generating disordered structures from target wave properties: stochastic[1,14], genetic[15], or topological[16] optimizations. However, traditional approaches are still challenging owing to the large design freedom inherited from disordered structures; thus, these approaches require very time-consuming and problem-specific processes to extract microstructural information at each stage of iterative and case-by-case design procedures. Until now, most works have focused on lower orders of microstructural descriptors (for example, two- or three-point probability functions) due to the significant complexity in calculating and interpreting higher-order descriptors[1]. However, even such simple descriptors have stimulated intriguing concepts and dynamics for disordered structures, such as hyperuniformity[17,18] for disordered bandgap materials[19].

To substitute the time-consuming and problem-specific process of calculating microstructural descriptors while making full use of microstructural information, we can envisage the use of multiple-layer neural network (NN) models to identify the relationship between disordered structures and wave behaviours. This deep-learning-based framework[20,21], one of the powerful machine learning (ML) tools, has proven successful for abstracting the features of datasets in pattern recognition, decision making, and language translation[22,23] when carefully preprocessed data can be used. Because of its applicability to general-purpose data formats, deep learning has recently been extended to handle a number of physics problems[24,25], such as classifications of crystals[26] or topological order[27], phase transitions and order parameters[28-30], optical device designs[31-36], and image reconstructions[37]. When we consider the vast amount of design freedom in disordered systems, deep learning will compose a powerful toolkit for resolving complexities in wave behaviours inside disordered structures, as shown in the inference of phases of matter using eigenfunctions[25,30].

Here, we employ deep convolutional neural networks (CNNs)[38] to identify the physical relationships between disordered structures and wave localization. The prediction of localization properties in disordered structures and the generation of necessary structures for target localizations are achieved with disorder-to-localization (D2L) and localization-to-disorder (L2D)

CNNs, respectively, by transforming disordered structures to multicolour images. Using dropout[39] or L2 regularization[21] techniques, the CNNs implemented with Google TensorFlow[40] are successfully trained with the expanded training dataset of collective and individual lattice deformations, even drawing an extrapolatory inference for the untrained regimes of disorder. Most importantly, our CNNs generate disordered structures with scale invariance following the power law, achieving an increase of two to four orders of magnitude in robustness to unexpected structural errors. We show that this ML-generated "scale-free" material with hub atoms inherits the properties of robustness to accidental attacks (or defects) and relative fragility to targeted attacks (or modulations)[41], in contrast to the "democratic" robustness of conventional normal-random disordered structures. The proposed approach can be applied to discover unexplored regimes of disorder in general wave systems and paves the way towards the design of materials by manipulating the ML architecture or the training process of ML network structures.

**Imaging disorder and localization**

We consider disordered structures obtained from the random deformation of a finite-size, two-dimensional (2D) square lattice of identical atoms (from Fig. 1a to 1b). Each atomic site of the lattice can describe a quantum-mechanical wavefunction of an atom[42], a phononic resonance of a metamaterial[43], or a propagating mode of an optical waveguide[44]. The standard tight-binding Hamiltonian of an $N$-atomic system governed by the eigenvalue equation $H\Psi_m = E_m\Psi_m$ ($m = 0$, 1, …, $N - 1$) is

$$H = \sum_i \varepsilon a_i^\dagger a_i + \sum_{i,j} \left( t_{ij} a_i^\dagger a_j + h.c. \right),$$ (1)

where $\varepsilon$ is the on-site energy, $a_i^\dagger$ (or $a_i$) is the creation (or annihilation) operator in the $i^{\text{th}}$ lattice site, $t_{ij}$ is the random hopping integral between the $i^{\text{th}}$ and $j^{\text{th}}$ lattice sites ($1 \leq i, j \leq N$), and $h.c.$

denotes the Hermitian conjugate. The disordered pattern is described by $t_{ij}$, which is determined by the spatial distance $d_{ij}$ between the $i^{th}$ and $j^{th}$ lattice sites. For generality, we consider all orders of hopping between lattice sites by defining the near-field hopping condition $t_{ij} = t_0 \exp(-\alpha d_{ij})$, where the coefficients $t_0$ and $\alpha$ are determined by an individual atomic Wannier function[45]. The distance $d_{ij}$ is adjusted by the perturbation on the position of each atom site (see Eq. (5) in Methods).

To develop D2L and L2D CNNs for the inference of wave-matter interactions, we devise a multicolour image representation of a disordered structure to be used as the CNN input. In this scenario, a 2D random displacement of an atomic site is projected along $x$ and $y$ spatial axes (Fig. 1c), and the resulting two ($x$ and $y$) projected layers from the entire disordered structure are assigned as two-colour images for CNNs (Figs. 1d and 1e). This projection can be directly extended into a 3D disordered structure, which leads to the sets of three-colour images with a tensor form.

The localization property of the proposed structure is quantified by the normalized mode area[46] $w_m$, which is defined by the inverse of the inverse participation ratio (IPR) as

$$w_m = \frac{1}{N} \frac{\left[ \sum_{s=1}^{N} \left( \psi_m{}^s \right)^2 \right]^2}{\sum_{s=1}^{N} \left( \psi_m{}^s \right)^4}, \tag{2}$$

where $\psi_m{}^s$ denotes the $s^{th}$ component of the eigenstate $\Psi_m$ ($s = 1, 2, \ldots, N$). The operation of the CNNs will then be the inference of the relationships between two-colour images (disordered structures) and a 1D array (mode area). The 1D mode area array is reshaped into a single-colour 2D image when it is used as the input to the L2D CNN, as discussed in Fig. 3.

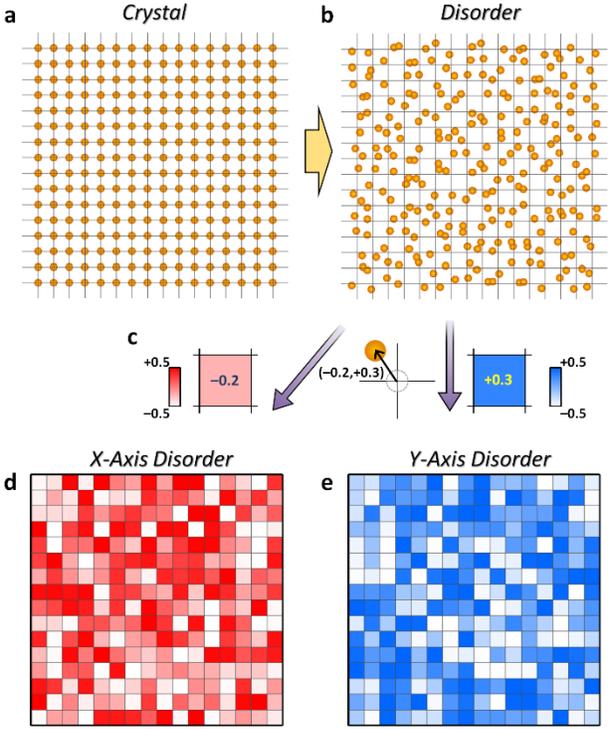

**Fig. 1. Multicolour image representation of disordered structures. a,** A 2D square lattice crystal and **b,** its deformation that generates a disordered structure. **c,** The projections of the 2D displacement of each atomic site along the *x*- and *y*-axes, which define the pixel values of the *x*-axis and *y*-axis colour images, respectively. **d, e,** The resulting two-colour images obtained from the disordered structure in **b**: **d,** the red-to-white image for the *x*-axis projection and **e,** the blue-to-white image for the *y*-axis projection.

### Disorder-to-Localization CNN

Figure 2a shows the network structure of the D2L CNN. For the two-colour image input, the CNN is composed of 3 cascaded convolution-pooling stages and the fully connected (FC) layer in front of the *N*-neuron output layer for the 1D array of $w_m$ (see Methods for network parameters). Each convolution-pooling stage is a series of the convolution (Conv) layer with 3×3 filters to extract a feature map and the max pooling layer to reduce the feature map size[20,21,38]. Because each mode has different degrees of localization, it is necessary to fairly estimate the regression error for a wide range of $w_m$ values. We thus employ the mean absolute percentage error (MAPE) as the cost function, which has been widely applied to regression and machine learning for forecasting models[47,48]. The MAPE cost function for the D2L CNN is expressed as

$$L_{\text{D2L}} = \sum_{m=0}^{N-1} \frac{\left| w_m^{\text{True}} - w_m^{\text{ML}} \right|}{w_m^{\text{True}}}, \tag{3}$$

where $w_m^{\text{ML}}$ is the D2L-CNN-calculated mode area and $w_m^{\text{True}}$ is the ground-truth mode area calculated by the Hamiltonian $H$ in Eq. (1).

The CNN is trained with the training dataset of randomly deformed lattices and their localization properties. The expanded training sets of $2 \times 10^4$ realizations are obtained by introducing both collective and individual deformations of atomic sites to improve the inference ability of the CNN (see Methods and Supplementary Note S1 for details of the deformation process). The validation accuracy of the CNN defined by $1 - L_{\text{D2L}}$ is monitored with the validation dataset of $1 \times 10^4$ realizations during the training. After training with the error backpropagation method[49], we calculate the test accuracy $1 - L_{\text{D2L}}$ of the trained CNN with the test dataset of $1 \times 10^4$ realizations (see Methods and Supplementary Note S2 for the training process). To monitor overfitting during and after the training, different random seeds for the deformation have been used in the training, validation, and test datasets.

Through the training process, we successfully trained the D2L CNN to predict disorder-induced localization. Figure 2b-e shows the ground-truth and ML prediction of the mode areas $w_m$ from given disordered structures: nearly crystallized (or weak disorder) (Figs. 2b and 2d) and nearly random (or strong disorder) (Figs. 2c and 2e) structures. We also compare the ground-truth (Fig. 2f) and ML-predicted (Fig. 2g) localization for a wide range of localization values of the test dataset ($1 \times 10^4$ realizations). Figures 2f and 2g are obtained by plotting $w_m$ of each realization as a function of the mode number $m$ and colouring each point according to the average mode area $w_{\text{avg}} = \sum w_m / N$ of each realization. We note that the ground-truth and ML-predicted localization shows excellent agreement for different values of $\underline{w}_{\text{avg}}$, achieving the test accuracy $1 - L_{\text{D2L}} \sim 94.80\%$. The trained D2L CNN enables an almost instantaneous prediction of localization properties for each mode from a given disordered structure without solving the eigenvalue problem of the Hamiltonian $H$ in Eq. (1).

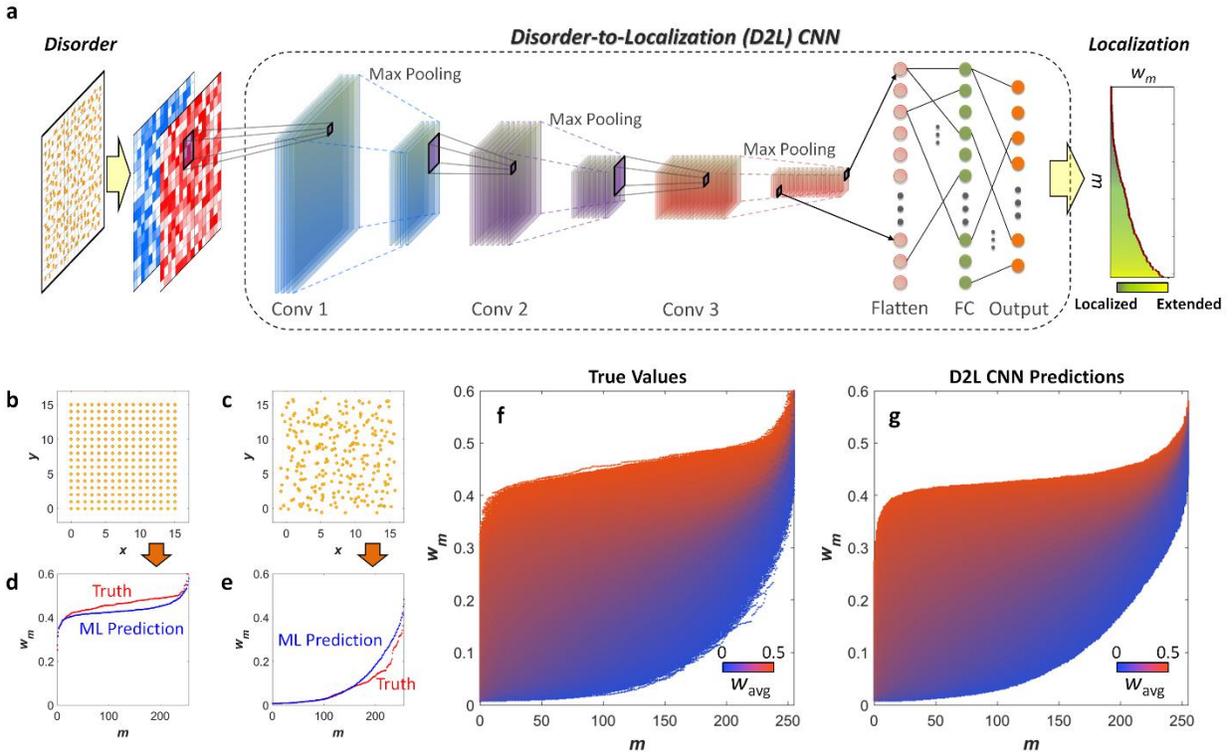

**Fig. 2. D2L CNN for predicting localization. a,** The network structure of the D2L CNN. The details of the network parameters are shown in the Methods section. **b-e,** The prediction of localization properties $w_m$ for **b, d** weakly disordered and **c, e** strongly disordered structures. **f, g,** Comparison of localizations between **f,** the ground truth and **g,** the D2L CNN prediction for a broad range of average mode area $w_{\text{avg}}$ of the test dataset ($1\times10^4$ realizations). The period of the $16\times16$ unperturbed square lattice (256 atoms) is set to 1 and the hopping parameters are $t_0 = 3.14 \times 10^{-2}$ and $\alpha = 1.1454$ throughout the manuscript. Although the on-site energy does not affect mode areas, we set the on-site energy as $\varepsilon = 1$ for energy spectra in the later discussion

(Supplementary Note S5). Mode numbers $m$ are sorted according to localization values in all examples throughout the manuscript.

**Localization-to-Disorder CNN**

As demonstrated in a classic question[50] of "Can one hear the shape of a drum?" and its answer[51], the relationship between a wave property (such as the localization or eigenspectrum) and material (or structural) platforms is non-unique, allowing multiple possible structures for a given wave property. This one-to-many relationship between a wave property and matter has made it difficult to achieve a stable inverse design of material from a given wave property because the existence of many solutions (matter) for an input (wave property) prohibits the stable convergence of the optimization for a cost function. In the inverse design of material using the ML method, several different approaches have been proposed to resolve this non-uniqueness problem: training of the input through a trained NN[31], training of the inverse NN from a trained forward NN[32,33], reinforcement learning[34], and iterative design of multiple NNs for each family of material structures with a given scattering property[35]. Considering the large design freedom in disordered structures, we employ the second approach[32,33]: training of the inverse L2D CNN using the pre-trained forward D2L CNN.

Figure 3a shows the network structure of the L2D CNN. The L2D CNN has the same network configuration as the D2L CNN (3 convolution-pooling stages and the FC layer), except for the input and output layer (see Methods for network parameters). The results of the L2D CNN from the $2N$ output neurons are reshaped to the two-colour images that represent the spatial profile of the ML-generated disordered structure. To guarantee the physical reality of the obtained solution, we utilize the "trained" D2L CNN with the fixed weight and bias parameters, which instantaneously predicts the localization in ML-generated disordered structures. The connection of the L2D CNN with the trained D2L CNN constructs the localization-to-disorder-to-localization (L2D2L) network (Fig. 3b), which effectively operates as the autoencoder for localization data. The MAPE cost function of the L2D2L CNN is defined as

$$L_{\text{L2D2L}} = \sum_{m=0}^{N-1} \frac{\left| w_m^{\text{Target}} - w_m^{\text{ML}} \right|}{w_m^{\text{Target}}}, \tag{4}$$

where $w_m^{\text{ML}}$ is the mode area calculated by the L2D2L CNN and $w_m^{\text{Target}}$ is the target mode area. The training of the entire L2D2L CNN (i.e., the partial training of the L2D CNN part) then allows the generation of disordered structures for the target wave localization (see Methods and Supplementary Note S2 for the training process). Training, validation, and test datasets are again prepared with different random seeds. We note that although the training dataset for the L2D2L CNN consists of localization data obtained from the tight-binding Hamiltonian in Eq. (1), the microstructural information used for the target localization data is not applied to the training of the L2D2L CNN.

The trained L2D CNN achieves a high test accuracy of $1 - L_{\text{L2D2L}} \sim 94.21\%$. We compare the target localizations (Fig. 3c) to the ML-predicted localizations obtained through the L2D2L CNN (Fig. 3d) and the Hamiltonian-calculated true values of the disordered structures generated by the L2D CNN (Fig. 3e), using the same data plotting format with those in Fig. 2f,g. Despite the good agreement between the target and true values (~79.10% between Figs. 3c and 3e), a non-negligible discrepancy exists near the strong localization regime with large deformations of atomic sites. We note that this accuracy degradation originates from the emergence of large deformations in the L2D-CNN-generated structure, which easily exceed the maximum deformation value inside the training datasets for the D2L CNN. Therefore, the test accuracy of the L2D CNN is restricted by the limit of the "extrapolation": the inference of the untrained regime of localization. The current good extrapolation could be further improved by expanding the range and type of training datasets and the number of hidden layers. However, we emphasize that large deformations themselves unveil for the first time a very intriguing but little recognized property in ML inverse designs[31-36]: the effect of the ML network structure on the ML-generated real-space structure, which enables the identification of scale-free properties for waves, as discussed in the later sections.

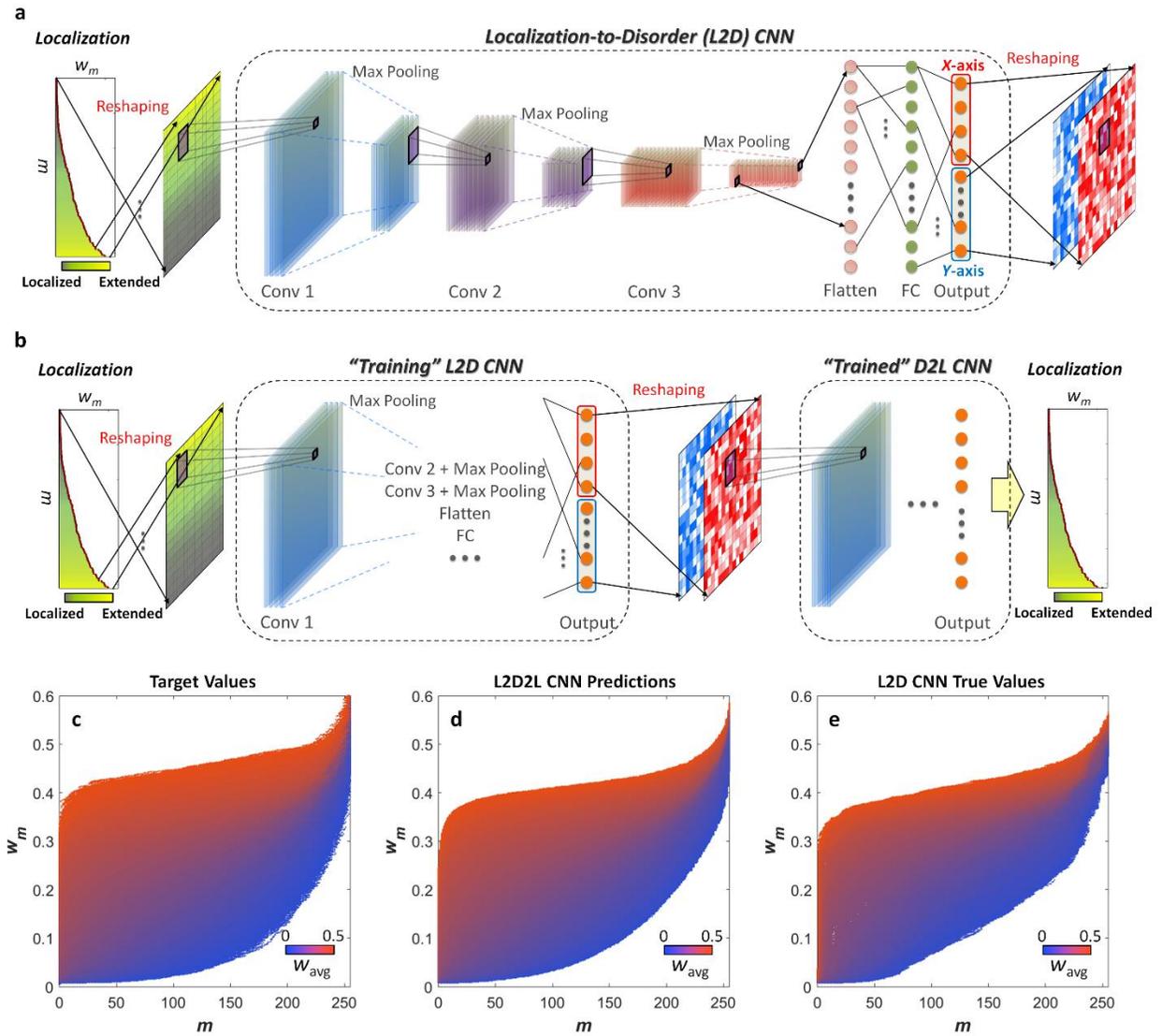

**Fig. 3. L2D CNN for generating disordered structures. a,** The network structure of the L2D CNN. **b,** The network structure of the L2D2L CNN for training the L2D CNN with the pre-

trained D2L CNN. The details of the network parameters are shown in the Methods section. **c-e,** Comparisons of localizations between **c,** the target values, **d,** the ML-predicted values from the L2D2L CNN, and **e,** the Hamiltonian-calculated true values with the disordered structures generated by the L2D CNN for a broad range of average mode area $w_{avg}$ of the test dataset ($1 \times 10^4$ realizations).

## Scale invariance in ML-generated microstructures

Due to the one-to-many relationship between a wave property and matter, the obtained ML-generated disordered structure corresponds to only one realization among numerous possible options for the target wave property. To examine the property of this ML "identification", in Fig. 4a-f, we compare the ML-generated structure with a seed structure having very similar localization properties. For the regimes of weak (Fig. 4a-c) and strong (Fig. 4d-f) disorder, we use initial seed structures (Figs. 4a and 4d) to obtain the target localization (red curves in Figs. 4c and 4f). By employing this target localization as an input of the trained L2D CNN, we achieve the corresponding ML-generated structures (Figs. 4b and 4e), which represent localization properties that are very similar to those of seed structures (black curves in Figs. 4c and 4f). However, surprisingly, the ML-generated structures consist of lattice deformations that are evidently different from the original deformations in the seed structures. This result originates from the training process of the L2D CNN, which is achieved from the training of the L2D2L CNN using only localization data (Fig. 3a) without the data of seed microstructures. The identification of the microstructure from the target localization can then have many possible options and is determined by the network structure of the L2D CNN, as discussed in Fig. 5.

For a deeper understanding of the differences between seed and ML-generated structures, we analyse the microstructural statistics of disordered structures by counting the distributions of the atomic site deformation $\Delta r_i = [(\Delta r_i^x)^2 + (\Delta r_i^y)^2]^{1/2}$, where $\Delta r_i^x$ and $\Delta r_i^y$ are the displacements of the $i^{th}$ atom along the $x$- and $y$-axes, respectively ($1 \leq i \leq N$; see Eq. (5) in Methods for seed structures, whereas $\Delta r_i$ of ML-generated structures is obtained from the L2D CNN). Figure 4g shows the microstructural statistics of the seed and ML-generated structures for 3200 realizations where the ML-generated structures have an average mode area $w_{avg}$ in the range of $0.20 \leq w_{avg} \leq 0.30$. We note that the seed and ML-generated structures show apparently differentiated statistics. First, the microstructural statistics of the seed structures follows a normal distribution due to the definition of Eq. (5) in Methods. However, as shown in the analysis based on the maximum-likelihood fitting method with goodness-of-fit tests[52,53] (inset (g-1) of Fig. 4g), the ML-generated class follows power-law statistics $(\Delta r)^{-\alpha}$ (inset (g-1) of Fig. 4g) and possesses a "heavy-tail" distribution (inset (g-2) of Fig. 4g). To guarantee the reliability of the power-law fitting result, in Supplementary Note S3, we analyse the power-law exponent $\alpha$ and the lower bound of the heavy tail $\Delta r_{min}$ for a different number of realizations. The result shows that the unique statistical distribution of ML-generated structures is maintained for a small number of realizations, from ~$10^1$ (2560 atoms) to ~$10^2$ (25600 atoms) realizations, and even a single realization also provides a similar value of $\alpha$ and $\Delta r_{min}$.

The result in Fig. 4g demonstrates that ML-generated disordered structures are composed of "scale-invariant" deformation without the characteristic perturbation strength of $\Delta r$. This finding is in sharp contrast to the characteristic $\Delta r$ of seed disordered structures, which is defined as the statistical centre of their normal distribution. We also note that the scale invariance of ML-generated disordered structures is universally observed for varying degrees of localization (Supplementary Note S4), which strongly implies that the identification of scale-invariant disordered structures originates from the properties of the L2D CNN, not from the observed wave-matter interactions.

Furthermore, the seed and ML-generated structures show very similar localization properties and distinct energy spectra (see Supplementary Note S5 for energy spectra). Therefore, the L2D CNN enables the independent and systematic handling of a part of wave quantities: here, the conservation of localization with an altered energy spectrum through the transformation of microstructural statics from normal-random to scale-invariant distributions. On the other side, among various possible realizations of disordered structures for a given wave property (here, localization) due to the one-to-many relationship between a wave and matter, the L2D CNN successfully selects one particular realization, which notably has the scale invariance in the structural profile.

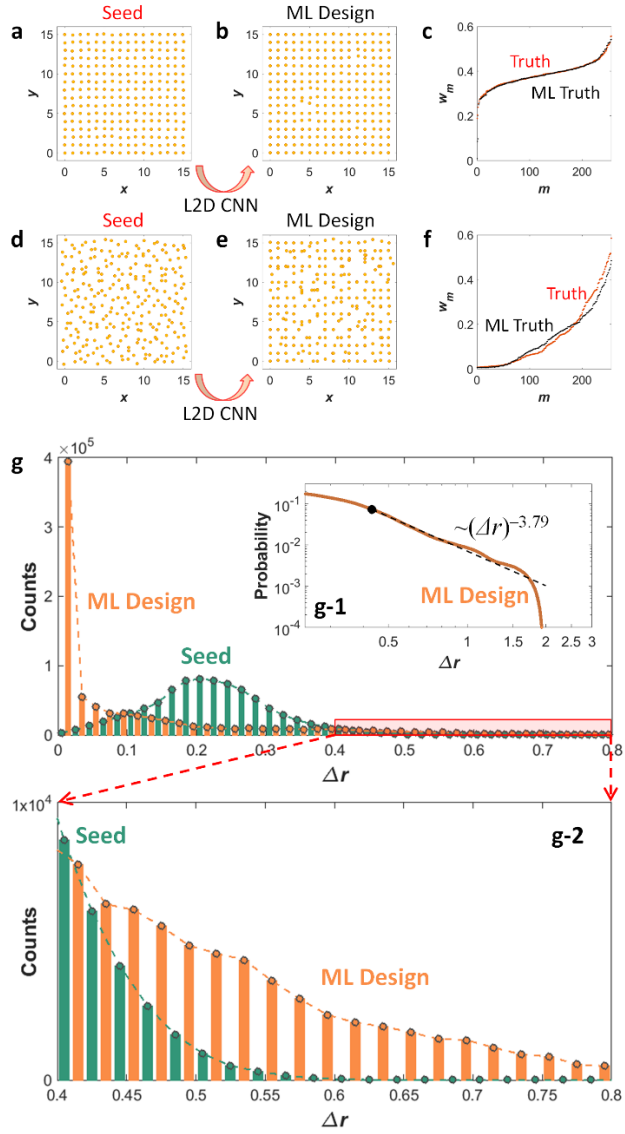

**Fig. 4. Scale-invariant disordered structures generated by the L2D CNN**. **a-f,** Comparison between seed and ML-generated structures for **a-c,** weak and **d-f,** strong disorder. **a, d,** Seed structures that provide the target localizations for the L2D CNN. **b, e,** ML-generated disordered structures obtained from the L2D CNN. **c, f,** Localization of seed ("Truth", red dotted lines) and ML-generated ("ML Truth", black dotted lines) structures, obtained from the tight-binding Hamiltonian in Eq. (1). **g,** Statistical distributions of the strength of the lattice deformation $\Delta r$ in the seed (green) and ML-generated (orange) structures for 3200 realizations satisfying $0.20 \leq w_{\mathrm{avg}} \leq 0.30$ in the ML design. The first inset **g-1** shows the log-log plot of **g** for the ML design, illustrating the power-law distribution. The orange line (composed of discretized points) represents the complementary cumulative distribution function (CDF) obtained from the data set

in **g**. The black dashed line represents the best fit to the data using the method in refs. 52,53, showing the power-law fitting of $(\Delta r)^{-3.79}$. The black dot represents the lower bound $\Delta r_{\min} = 0.432$ to the power-law behaviour. The second inset **g-2** shows the extended plot of the range $0.4 \leq \Delta r \leq 0.8$, demonstrating the heavy tail of the distribution for the ML design.

Because the values of the output neurons in the L2D CNN determines the lattice deformation in ML-generated structures, the scale invariance in the deformation is strongly related to the ML network structure (weight and bias distributions) of the L2D CNN. To examine this conjecture, in Fig. 5, we analyse the relationship between the microstructural statistics of ML-generated structures and the network structure of the L2D CNN. Among numerous weight and bias parameters ($\sim 1.5 \times 10^7$ parameters each in the D2L and L2D CNNs), the most critical parameters are the weights from the FC layer (2048 neurons) to the output layer (512 neurons) in the L2D CNN, which are described by a 2048×512 matrix. Although the weights and bias in hidden layers should also affect the output layer neurons indirectly, we expect that this indirect effect is less significant than the direct effect from the FC-output weights.

For $w_{ji}^x$ and $w_{ji}^y$, which denote the weights from the $i^{\text{th}}$ FC neuron to the $j^{\text{th}}$ $x$-axis and $y$-axis output neurons, respectively ($1 \leq i \leq 2048$ and $1 \leq j \leq 256$ in our design), we define the strength of the weights to the $j^{\text{th}}$ output neuron (or the $j^{\text{th}}$ atom in an ML-generated disordered structure) as $W_j = \sum_i [(w_{ji}^x)^2 + (w_{ji}^y)^2]$. Figure 5b shows the CDF of $W_j$, which represents a very similar statistical distribution with $\Delta r$ in terms of its inflection point (Fig. 5a) and also possesses the heavy-tailed distribution. The comparison between Figs. 5a and 5b provides clear-cut evidence of the effect of the ML network structure on ML-generated materials. This finding becomes more evident by examining different ML architectures which lead to different weights and bias distributions. In Supplementary Note S6, we investigate another D2L and L2D CNN each with a single pooling stage, which enables the control of the $W_j$ distribution and the following alteration of ML-generated structures. We note that the heavy-tailed distribution is also maintained in this single-pooling-layer design.

To guarantee the generality of the observed scale-free properties, we also examine the effect of the test accuracy on the scale invariance (Fig. 5c,d). Among 3200 realizations in the example in Fig. 4, we select the sets of ML-generated structures having high ($\geq$ 84%, 194 realizations, Fig. 5c) and low ($\leq$ 69%, 191 realizations, Fig. 5d) test accuracies. We note that both cases possess very similar statistical distributions with the power-law fitting result. This result again confirms that the scale invariance originates from the statistical distribution of the ML architecture, not from the mismatch between the ML result and theoretical truth.

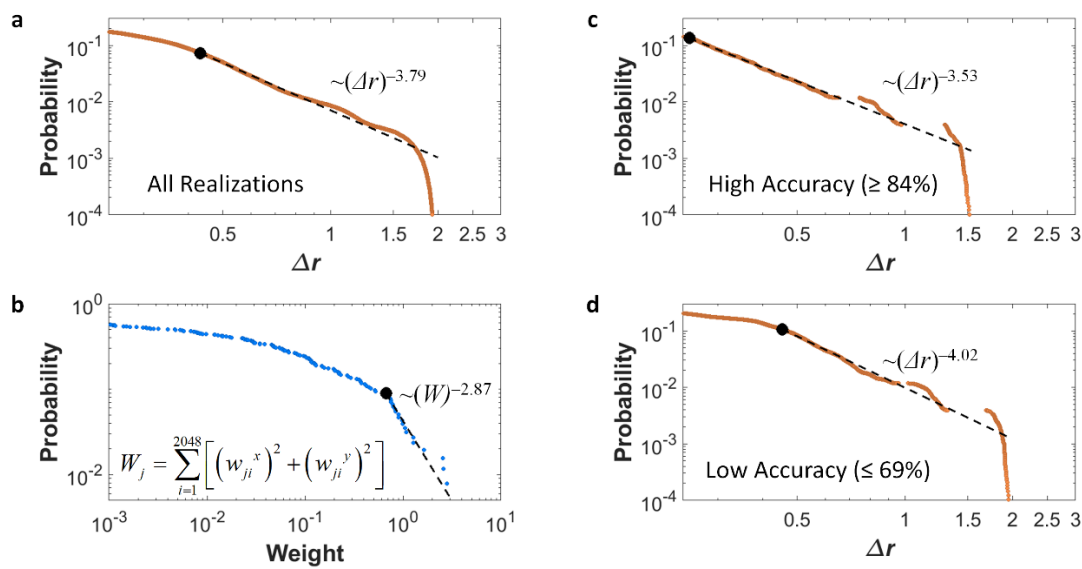

**Fig. 5. Relationships between ML-generated disordered structures, ML architecture, and test accuracies in terms of scale invariance**. **a,** Power-law fitting of the statistical distribution of $\Delta r$ in ML-generated disordered structures, which is the same figure with Fig. 4g-1 and is shown for comparison. **b,** Power-law fitting of the statistical distribution of the weight strength parameter $W_j$. **c, d,** Power-law fitting results of the realizations for **c,** high ($\geq$ 84%) and **d,** low ($\geq$ 69%) test accuracies. All of the fitting results are based on the same method using in Fig. 4g-1.

**Scale-free materials with hub atoms**

The scale invariance in microstructural statistics (Figs. 4 and 5) imposes intriguing characteristics on ML-generated disordered structures: "scale-free" properties on waves. A scale-free properties, which represent the power-law probabilistic distribution with heavy-tailed statistics, has been one of the most influential concepts in network science[2,54], data science[52,53], and random matrix theory[55,56]. In addition to its ubiquitous nature in biological, social, and technological systems[2], the most important impact of scale-free property is the emergence of core nodes, also known as "hubs", which possess a very large number of links or interactions, thereby governing signal transport inside the system[2,41,54]. The existence of hub nodes strongly correlates with the robustness of scale-free systems: fault-tolerant behaviours, especially superior robustness to accidental attacks and relative fragility to targeted attacks[2,41,57].

Although the scale-free nature is well defined in the infinite-size limit[2,41,54], similar to the condition of ergodicity in random heterogeneous materials[1], the power-law microstructural statistics of our systems with the long-tail distribution lead to well-defined hub behaviours and the following robustness of wave properties. To investigate the robustness of our wave systems, we exert the "attack" (material imperfection, system error, or modulation) on each atom of disordered structures to adjust their localization properties. The attack is defined by the position perturbation of each atom as $r_i^x = r_i^{x0} + \rho_a \cos[u_i(0, 2\pi)]$ and $r_i^y = r_i^{y0} + \rho_a \sin[u_i(0, 2\pi)]$, where $r_i^{x,y}$ (or $r_i^{x0,y0}$) are the $x$ and $y$ perturbed (or original) positions of the $i^{\text{th}}$ atom in a disordered structure, $\rho_a$ is the perturbation strength, and $u_i(a,b)$ is the random value for the $i^{\text{th}}$ atom from the uniform random distribution between $a$ and $b$.

Figures 6a and 6b show the degree of robustness in two disordered structures with different microstructural statistics in terms of the perturbation of localization $\Delta w_m$. The attack is applied to each atom of normal-random seed ($w_{\mathrm{avg}} = 0.145$) and scale-free ML-generated ($w_{\mathrm{avg}} = 0.140$) disordered structures, which have similar localization properties (~84.05% test accuracy). Remarkably, compared with the seed structure, the scale-free disordered structure shows a reduction of two to four orders of magnitude in the perturbation of mode areas $\Delta w_m$, especially in highly localized modes (small $m$). This result demonstrates that the scale-free ML-generated disorder provides more robust localization properties than the normal-random seed disorder, following fault-tolerant behaviours in general scale-free systems[2,41,57].

In Figs. 6c and 6d, we also demonstrate the existence of hub atoms, which is the origin of the robustness of scale-free systems[2,41]. To detect "hub atoms" in disordered structures, we define the normalized error $\delta$ that measures the average perturbation of the mode area $\Delta w_m$ obtained by attacking a specific atom. First, the apparent "democratic" response of $\delta$, which represents the nearly equal perturbation of $\Delta w_m$ regardless of the perturbed atom position, is observed in the normal-random seed structure (Fig. 6c), following the signal behaviour in Erdős-Rényi random systems[2,58]. In contrast, our ML-generated scale-free disordered structure is no longer democratic; some "hub" atoms derive more sensitive responses (larger $\delta$) to the perturbation (Fig. 6d), following the signal behaviour in Barabási-Albert scale-free systems[2,41]. This result successfully demonstrates the scale-free nature of our ML-generated disorder: highly robust localization to accidental perturbations and relatively fragile localization to targeted perturbations on hub atomic sites.

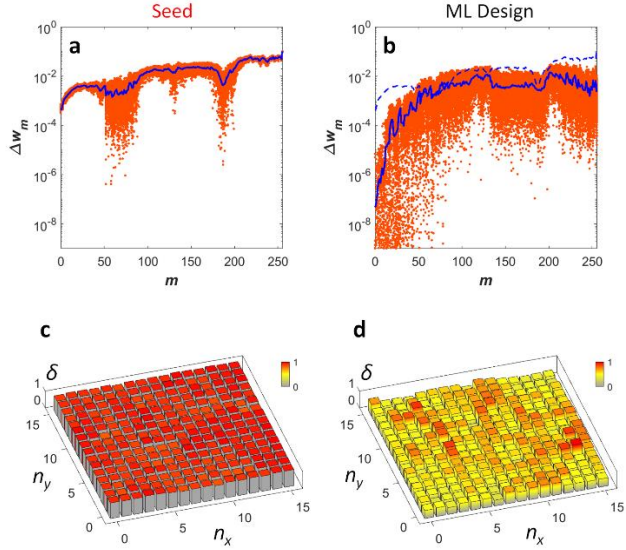

**Fig. 6. Scale-free properties of ML-generated disorder with hub atoms**. **a, b,** Comparison of the robustness in **a,** seed and **b,** ML-generated disordered structures in terms of the perturbation in the mode area from the attack (or error) to a specific atom. Each red point denotes the perturbation of the $m^{\text{th}}$ mode area $\varDelta w_m$ by imposing the attack to a specific atom. Blue solid lines represent the average perturbation. The blue dashed line in **b** is the average perturbation of the seed structure shown in **a** for comparison. **c, d,** Normalized errors $\delta$ for attacking each atom in the **c,** seed and **d,** ML-generated disordered structures. Larger $\delta$ denotes a more sensitive response of wave localization to the attack. $n_x$ and $n_y$ denote the $x$ and $y$ indices in the unperturbed square lattice, respectively ($0 \leq n_x, n_y \leq 15$ for 256 atoms).

## Discussion

In conclusion, we demonstrated for the first time that the ML approach can identify disordered materials with the target localization, which also have scale-free properties for waves. Instead of calculating microstructural descriptors for analysing disordered structures, we proposed a CNN-based modelling approach for wave-matter interactions, by using convolution processes in CNNs to abstract and map the relationship between localization and disordered structures. With successful training results for the ML prediction and generation of wave-matter interactions, we showed that ML-generated disordered structures possess scale invariance with power-law microstructural statistics, which allows the realization of scale-free materials for waves with excellent robustness in terms of wave behaviours and hub dynamics.

We also demonstrated that the lattice deformation is strongly related to the weights of the output neurons in the L2D CNN, showing the similar heavy-tailed distributions. In this context, the apparent stochastic difference between normal-random seed structures and scale-free L2D CNN outputs raises an interesting open question; the training process of deep NNs could inherently possess the scale-free property. Recently, in random matrix theory, it was demonstrated that the correlations in the weight matrices of well-trained deep NNs can be fit to a power law with the heavy-tailed distribution[55,56]. This theory enables the successful analogy between ML network structures and ML-generated real-space wave structures in our result: the identification of the "heavy-tailed perturbation distribution" of atomic sites using the "heavy-tailed weight distribution" of CNN neurons. While these complex systems in software and real space emphasize the role of the "heavy tail" in the statistical distribution, the optimization process of the CNNs in this viewpoint corresponds to the evolutionary process of realizing general scale-free systems[2,41,54]. We also note that exploring ML architectures to control scale-free properties or even realize non-scale-free distributions will inspire exciting future research in material science and wave physics. For the inverse design of disordered systems and the following statistical analysis of ML-generated materials in terms of scale-free properties, the applications of reinforcement learning, unsupervised learning, or well-trained ML networks such as U-net[59,60] would also be an excellent topic for study.

Scale-free materials discovered by the ML method will stimulate a new design strategy for general wave devices in disordered structures, such as lasing[12,61], energy storage[62], and complete bandgap materials[19]. Scale invariance can significantly improve the performance of these wave devices by achieving robustness to accidental errors (such as unwanted defects in fabrications or measurements) and the fragility to targeted errors (such as the intended system modulation for active devices). Along with the ML generation of scale-free structures with target wave properties, our results will motivate further research on controlling CNN training or selecting different CNN architectures, which will enable the generation of wave structures analogous to various types of complex systems, such as small-world, modular, or self-similar systems. The obtained scale-free wave material will also offer new insight into other scale-free-type material structures, such as Lévy glasses with superdiffusion[63,64]: the microstructural realization of a random walk having step lengths with a power-law distribution.

**Methods**

**Network structures and training hyperparameters of D2L and L2D CNNs.** For $N = 16 \times 16$ atomic lattices, the D2L CNN accepts two $16 \times 16$ images as the input (a disordered structure), whereas the L2D CNN accepts a single $16 \times 16$ image as the input (a reshaped mode area). For both D2L and L2D CNNs, the numbers of filters (or the thicknesses) of the convolution layers are set to 256, 512, and 1024 in the first, second, and third layers, respectively. We use zero padding to maintain the spatial dimensions of feature maps during the convolution processes[21,38]. The max pooling layer leads to the down-sampling of feature maps by extracting the maximum value of each patch with a stride of 2 pixels[20,21]. The result of 3 cascaded convolution-pooling states is reshaped (or flattened) to a 1D array and is then connected to the FC layer, which has 2048 neurons. The FC layer is connected to the $N$-atomic output layer in the D2L CNN for the mode area $w_m$ and is connected to the $2N$-atomic output layer in the L2D CNN for two-colour images that describe a disordered structure.

To avoid a vanishing gradient problem during training, we use the rectified linear unit (ReLU) activation for each layer of CNNs. We utilize the Adam optimization function[49] with exponential decay in the learning rate for stable convergence and employ a mini-batch of size 10 for efficient learning. To avoid overfitting, we apply the dropout method[39] in the D2L CNN by randomly keeping 50% of neurons in the FC layer during training and apply the L2 regularization[21] in the L2D CNN (TensorFlow scale parameter: 0.05) to suppress excessively large values of weights. The learning processes of the D2L and L2D CNNs are shown in Supplementary Note S2. All ML computations were performed on a single desktop computer with two NVIDIA GeForce RTX 2080 Ti GPUs.

**Deformation of lattices for datasets.** To train the CNNs, avoiding overfitting to a certain type of disordered structure, the carefully preprocessed training dataset has to cover a wide range of the relationship between disordered structures and localization from large to small values of $w_{avg}$ $= \sum w_m / N$. For this purpose, we assign the collective and individual deformations of atomic sites as

$$\Delta r_i^x = \rho \cos\left(u_i\left(0, 2\pi\right)\right) + u_i\left(-\sigma, +\sigma\right),$$
$$\Delta r_i^y = \rho \sin\left(u_i\left(0, 2\pi\right)\right) + u_i\left(-\sigma, +\sigma\right),$$

(5)

where $\Delta r_i^x$ and $\Delta r_i^y$ denote the displacements of the $i^{th}$ atom along the $x$- and $y$-axes ($1 \leq i \leq N$), respectively; $u_i(a,b)$ is the random value for the $i^{th}$ atom from the uniform random distribution between $a$ and $b$; $\rho$ is the amplitude of the collective displacement of all atoms; and $\sigma$ is the amplitude of the individual displacement of each atom. The strengths of the collective and individual deformations are randomly assigned for each realization of the dataset, as $\rho = \rho_{max}u(0,$

1) and $\sigma = \sigma_{max}u(0, 1)$, where $u(a,b)$ is the random value assigned to each realization from the uniform random distribution between $a$ and $b$. We set $\rho_{max} = 0.6$ and $\sigma_{max} = 0.6$ for all examples in this manuscript. The comparison between collective and individual deformations through different values of $\rho_{max}$ and $\sigma_{max}$ are shown in Supplementary Note S1.

**Data availability**

The data that support the plots and other findings of this study are available from the corresponding author upon request.

**Code availability**

All code developed in this work will be made available from the corresponding author upon request.

**References**

1    Torquato, S. *Random heterogeneous materials: microstructure and macroscopic properties*. Vol. 16 (Springer Science & Business Media, 2002).

2    Barabási, A.-L. *Network science*. (Cambridge university press, 2016).

3    Wiersma, D. S. Disordered photonics. *Nat. Photon.* **7**, 188-196 (2013).

4    Anderson, P. W. Absence of diffusion in certain random lattices. *Phys. Rev.* **109**, 1492 (1958).

5    Van Albada, M. P. & Lagendijk, A. Observation of weak localization of light in a random medium. *Phys. Rev. Lett.* **55**, 2692 (1985).

6    Jiang, X., Shao, L., Zhang, S.-X., Yi, X., Wiersig, J., Wang, L., Gong, Q., Lončar, M.,

Yang, L. & Xiao, Y.-F. Chaos-assisted broadband momentum transformation in optical microresonators. *Science* **358**, 344-347 (2017).

7      Hsu, C. W., Goetschy, A., Bromberg, Y., Stone, A. D. & Cao, H. Broadband coherent enhancement of transmission and absorption in disordered media. *Phys. Rev. Lett.* **115**, 223901 (2015).

8      Stützer, S., Plotnik, Y., Lumer, Y., Titum, P., Lindner, N. H., Segev, M., Rechtsman, M. C. & Szameit, A. Photonic topological Anderson insulators. *Nature* **560**, 461 (2018).

9      Chabé, J., Lemarié, G., Grémaud, B., Delande, D., Szriftgiser, P. & Garreau, J. C. Experimental observation of the Anderson metal-insulator transition with atomic matter waves. *Phys. Rev. Lett.* **101**, 255702 (2008).

10     Wiersma, D. S., Bartolini, P., Lagendijk, A. & Righini, R. Localization of light in a disordered medium. *Nature* **390**, 671 (1997).

11     Segev, M., Silberberg, Y. & Christodoulides, D. N. Anderson localization of light. *Nat. Photon.* **7**, 197-204 (2013).

12     Liu, J., Garcia, P., Ek, S., Gregersen, N., Suhr, T., Schubert, M., Mørk, J., Stobbe, S. & Lodahl, P. Random nanolasing in the Anderson localized regime. *Nat. Nanotech.* **9**, 285-289 (2014).

13     Sheinfux, H. H., Lumer, Y., Ankonina, G., Genack, A. Z., Bartal, G. & Segev, M. Observation of Anderson localization in disordered nanophotonic structures. *Science* **356**, 953-956 (2017).

14     Yeong, C. & Torquato, S. Reconstructing random media. *Phys. Rev. E* **57**, 495 (1998).

15     Weber, T. & Bürgi, H.-B. Determination and refinement of disordered crystal structures using evolutionary algorithms in combination with Monte Carlo methods. *Acta*

*Crystallogr. A* **58**, 526-540 (2002).

16      Eschenauer, H. A. & Olhoff, N. Topology optimization of continuum structures: a review. *Appl. Mech. Rev.* **54**, 331-390 (2001).

17      Torquato, S. & Stillinger, F. H. Local density fluctuations, hyperuniformity, and order metrics. *Phys. Rev. E* **68**, 041113 (2003).

18      Torquato, S. Hyperuniform states of matter. *Phys. Rep.* **745**, 1 (2018).

19      Man, W., Florescu, M., Williamson, E. P., He, Y., Hashemizad, S. R., Leung, B. Y., Liner, D. R., Torquato, S., Chaikin, P. M. & Steinhardt, P. J. Isotropic band gaps and freeform waveguides observed in hyperuniform disordered photonic solids. *Proc. Natl. Acad. Sci. USA* **110**, 15886-15891 (2013).

20      LeCun, Y., Bengio, Y. & Hinton, G. Deep learning. *Nature* **521**, 436 (2015).

21      Goodfellow, I., Bengio, Y. & Courville, A. *Deep learning*. (MIT press, 2016).

22      Silver, D., Huang, A., Maddison, C. J., Guez, A., Sifre, L., Van Den Driessche, G., Schrittwieser, J., Antonoglou, I., Panneershelvam, V. & Lanctot, M. Mastering the game of Go with deep neural networks and tree search. *Nature* **529**, 484 (2016).

23      Mnih, V., Kavukcuoglu, K., Silver, D., Rusu, A. A., Veness, J., Bellemare, M. G., Graves, A., Riedmiller, M., Fidjeland, A. K. & Ostrovski, G. Human-level control through deep reinforcement learning. *Nature* **518**, 529 (2015).

24      Carleo, G., Cirac, I., Cranmer, K., Daudet, L., Schuld, M., Tishby, N., Vogt-Maranto, L. & Zdeborová, L. Machine learning and the physical sciences. *Rev. Mod. Phys.* **91**, 045002 (2019).

25      Ohtsuki, T. & Mano, T. Drawing Phase Diagrams of Random Quantum Systems by Deep Learning the Wave Functions. *J. Phys. Soc. Jpn.* **89**, 022001 (2020).

26    Ziletti, A., Kumar, D., Scheffler, M. & Ghiringhelli, L. M. Insightful classification of crystal structures using deep learning. *Nat. Commun.* **9**, 2775 (2018).

27    Rodriguez-Nieva, J. F. & Scheurer, M. S. Identifying topological order through unsupervised machine learning. *Nat. Phys.* **15**, 790-795 (2019).

28    Carrasquilla, J. & Melko, R. G. Machine learning phases of matter. *Nat. Phys.* **13**, 431 (2017).

29    Broecker, P., Carrasquilla, J., Melko, R. G. & Trebst, S. Machine learning quantum phases of matter beyond the fermion sign problem. *Sci. Rep.* **7**, 1-10 (2017).

30    Ohtsuki, T. & Ohtsuki, T. Deep learning the quantum phase transitions in random two-dimensional electron systems. *J. Phys. Soc. Jpn.* **85**, 123706 (2016).

31    Peurifoy, J., Shen, Y., Jing, L., Yang, Y., Cano-Renteria, F., DeLacy, B. G., Joannopoulos, J. D., Tegmark, M. & Soljačić, M. Nanophotonic particle simulation and inverse design using artificial neural networks. *Sci. Adv.* **4**, eaar4206 (2018).

32    Liu, Z., Zhu, D., Rodrigues, S. P., Lee, K.-T. & Cai, W. Generative model for the inverse design of metasurfaces. *Nano Lett.* **18**, 6570-6576 (2018).

33    Liu, D., Tan, Y., Khoram, E. & Yu, Z. Training deep neural networks for the inverse design of nanophotonic structures. *ACS Photon.* **5**, 1365-1369 (2018).

34    Sajedian, I., Badloe, T. & Rho, J. Optimisation of colour generation from dielectric nanostructures using reinforcement learning. *Opt. Express* **27**, 5874-5883 (2019).

35    Baxter, J., Lesina, A. C., Guay, J.-M., Weck, A., Berini, P. & Ramunno, L. Plasmonic colours predicted by deep learning. *Sci. Rep.* **9**, 8074 (2019).

36    Ma, W., Cheng, F. & Liu, Y. Deep-learning-enabled on-demand design of chiral metamaterials. *ACS Nano* **12**, 6326-6334 (2018).

37    Rivenson, Y., Zhang, Y., Günaydın, H., Teng, D. & Ozcan, A. Phase recovery and holographic image reconstruction using deep learning in neural networks. *Light Sci. Appl.* **7**, 17141 (2018).

38    Krizhevsky, A., Sutskever, I. & Hinton, G. E. *Imagenet classification with deep convolutional neural networks*. In *Advances in neural information processing systems* **25**, 1097-1105 (2012).

39    Srivastava, N., Hinton, G., Krizhevsky, A., Sutskever, I. & Salakhutdinov, R. Dropout: a simple way to prevent neural networks from overfitting. *J. Mach. Learn. Res.* **15**, 1929-1958 (2014).

40    Abadi, M., Agarwal, A., Barham, P., Brevdo, E., Chen, Z., Citro, C., Corrado, G. S., Davis, A., Dean, J. & Devin, M. Tensorflow: Large-scale machine learning on heterogeneous distributed systems. *arXiv preprint arXiv:1603.04467* (2016).

41    Barabási, A.-L. & Bonabeau, E. Scale-Free Networks. *Sci. Am.* **288**, 60-69 (2003).

42    Economou, E. & Antoniou, P. Localization and off-diagonal disorder. *Solid State Commun.* **21**, 285-288 (1977).

43    Serra-Garcia, M., Peri, V., Süsstrunk, R., Bilal, O. R., Larsen, T., Villanueva, L. G. & Huber, S. D. Observation of a phononic quadrupole topological insulator. *Nature* **555**, 342 (2018).

44    El Hassan, A., Kunst, F. K., Moritz, A., Andler, G., Bergholtz, E. J. & Bourennane, M. Corner states of light in photonic waveguides. *Nat. Photon.*, 1-4 (2019).

45    Ashcroft, N. W., Mermin, N. D. & Rodriguez, S. *Solid state physics*. (Cengage Learning, 1976).

46    Schwartz, T., Bartal, G., Fishman, S. & Segev, M. Transport and Anderson localization in disordered two-dimensional photonic lattices. *Nature* **446**, 52-55 (2007).

47    Yu, R., Li, Y., Shahabi, C., Demiryurek, U. & Liu, Y. *Deep learning: A generic approach for extreme condition traffic forecasting*. In *Proceedings of the 2017 SIAM international Conference on Data Mining* 777-785 (2017).

48    Yildiz, B., Bilbao, J. I. & Sproul, A. B. A review and analysis of regression and machine learning models on commercial building electricity load forecasting. *Renewable and Sustainable Energy Reviews* **73**, 1104-1122 (2017).

49    Kingma, D. P. & Ba, J. Adam: A method for stochastic optimization. *arXiv preprint arXiv:1412.6980* (2014).

50    Kac, M. Can one hear the shape of a drum? *Am. Math. Monthly* **73**, 1-23 (1966).

51    Gordon, C., Webb, D. L. & Wolpert, S. One cannot hear the shape of a drum. *Bull. Am. Math. Soc.* **27**, 134-138 (1992).

52    Clauset, A., Shalizi, C. R. & Newman, M. E. Power-law distributions in empirical data. *SIAM review* **51**, 661-703 (2009).

53    Alstott, J. & Bullmore, D. P. powerlaw: a Python package for analysis of heavy-tailed distributions. *PloS one* **9** (2014).

54    Barabási, A.-L. & Albert, R. Emergence of scaling in random networks. *Science* **286**, 509-512 (1999).

55    Martin, C. H. & Mahoney, M. W. Implicit self-regularization in deep neural networks: Evidence from random matrix theory and implications for learning. *arXiv preprint arXiv:1810.01075* (2018).

56    Martin, C. H. & Mahoney, M. W. Heavy-tailed Universality predicts trends in test accuracies for very large pre-trained deep neural networks. *arXiv preprint*

*arXiv:1901.08278* (2019).

57    Cohen, R., Erez, K., Ben-Avraham, D. & Havlin, S. Breakdown of the internet under intentional attack. *Phys. Rev. Lett.* **86**, 3682 (2001).

58    Erdős, P. & Rényi, A. On the evolution of random graphs. *Publ. Math. Inst. Hung. Acad. Sci* **5**, 17-60 (1960).

59    Ronneberger, O., Fischer, P. & Brox, T. *U-net: Convolutional networks for biomedical image segmentation*. In *International Conference on Medical image computing and computer-assisted intervention*  234-241 (2015).

60    Rosu, R. A., Schütt, P., Quenzel, J. & Behnke, S. Latticenet: Fast point cloud segmentation using permutohedral lattices. *arXiv preprint arXiv:1912.05905* (2019).

61    Bittner, S., Guazzotti, S., Zeng, Y., Hu, X., Yılmaz, H., Kim, K., Oh, S. S., Wang, Q. J., Hess, O. & Cao, H. Suppressing spatiotemporal lasing instabilities with wave-chaotic microcavities. *Science* **361**, 1225-1231 (2018).

62    Liu, C., Di Falco, A., Molinari, D., Khan, Y., Ooi, B. S., Krauss, T. F. & Fratalocchi, A. Enhanced energy storage in chaotic optical resonators. *Nat. Photon.* **7**, 473-478 (2013).

63    Bertolotti, J., Vynck, K., Pattelli, L., Barthelemy, P., Lepri, S. & Wiersma, D. S. Engineering disorder in superdiffusive Levy glasses. *Adv. Funct. Mater.* **20**, 965-968 (2010).

64    Burresi, M., Radhalakshmi, V., Savo, R., Bertolotti, J., Vynck, K. & Wiersma, D. S. Weak localization of light in superdiffusive random systems. *Phys. Rev. Lett.* **108**, 110604 (2012).

**Acknowledgements**

We acknowledge financial support from the National Research Foundation of Korea (NRF) through the Global Frontier Program (S.Y., X.P., N.P.: 2014M3A6B3063708), the Basic Science Research Program (S.Y.: 2016R1A6A3A04009723), and the Korea Research Fellowship Program (X.P., N.P.: 2016H1D3A1938069), all funded by the Korean government.

**Author contributions**

S.Y. conceived the idea presented in the manuscript. S.Y. and X.P. developed the theory and ML codes using Google TensorFlow. N.P. encouraged S.Y. and X.P. to investigate disordered systems for waves using ML and network theory while supervising the findings of this work. All authors discussed the results and contributed to the final manuscript.

**Competing interests**

The authors have no conflicts of interest to declare.

**Correspondence and requests for materials** should be addressed to N.P. (nkpark@snu.ac.kr).

**Supplementary Information for "Machine learning identifies scale-free properties in disordered materials"**

Sunkyu Yu, Xianji Piao, and Namkyoo Park[*]

Photonic Systems Laboratory, Department of Electrical and Computer Engineering, Seoul National University, Seoul 08826, Korea

*E-mail address for correspondence: nkpark@snu.ac.kr (N.P.)

**Note S1. Training datasets from collective and individual deformations**

**Note S2. Training process of the D2L and L2D CNNs**

**Note S3. Dependence of power-law distributions on data size**

**Note S4. Scale invariance in ML-generated materials with different degrees of localization**

**Note S5. Energy spectra in normal-random and scale-free disorder**

**Note S6. Scale invariance in a different ML architecture**

**Note S1. Training datasets from collective and individual deformations**

To avoid overfitting in machine learning (ML), it is helpful to prepare a good training dataset that includes representative examples for each class of features. Therefore, for the inference of the relationship between wave localization and disordered structures, we need to prepare a dataset that covers the maximum range of microstructural patterns and wave localization values.

For this purpose, we compare the collective and individual deformations of atomic sites, as discussed in the Methods section. First, $\rho_{max} \neq 0$ and $\sigma_{max} = 0$ results in the collective deformation: the same amount of perturbation for all atoms, while the randomly assigned azimuthal angle $u_i(0, 2\pi)$ to the $i^{th}$ atom leads to disordered structures with broken discrete translational symmetry, and $\rho = \rho_{max} u(0, 1)$ leads to different levels of perturbation for each realization. This collective deformation provides rigorously homogeneous patterns of perturbation strength. On the other hand, $\rho_{max} = 0$ and $\sigma_{max} \neq 0$ results in the individual deformation: randomly assigned perturbation for each atom, which has an average perturbation of $\sigma = \sigma_{max} u(0, 1)$ for each spatial axis. This individual deformation provides locally inhomogeneous but statistically homogeneous perturbation strength [1]. Figure S1a-d shows the localization and its statistics for collective (Figs. S1a and S1b) and individual (Figs. S1c and S1d) deformations. Although localization properties are similar in both deformations, their microstructural patterns of perturbation strength are different: collective patterns, which are homogeneous in all length scales (Figs. S1a and S1b), and individual patterns, which are locally inhomogeneous and statistically homogeneous (Figs. S1c and S1d). To cover the mixing of collective and individual deformations, we set the condition of $\rho_{max} \neq 0$ and $\sigma_{max} \neq 0$ (see Figs. S1e and S1f for the localization properties). The combination of collective and individual deformations provides more equally distributed localization values for the dataset (see dashed lines in Fig. S1f), which prevents overfitting to certain values of localization.

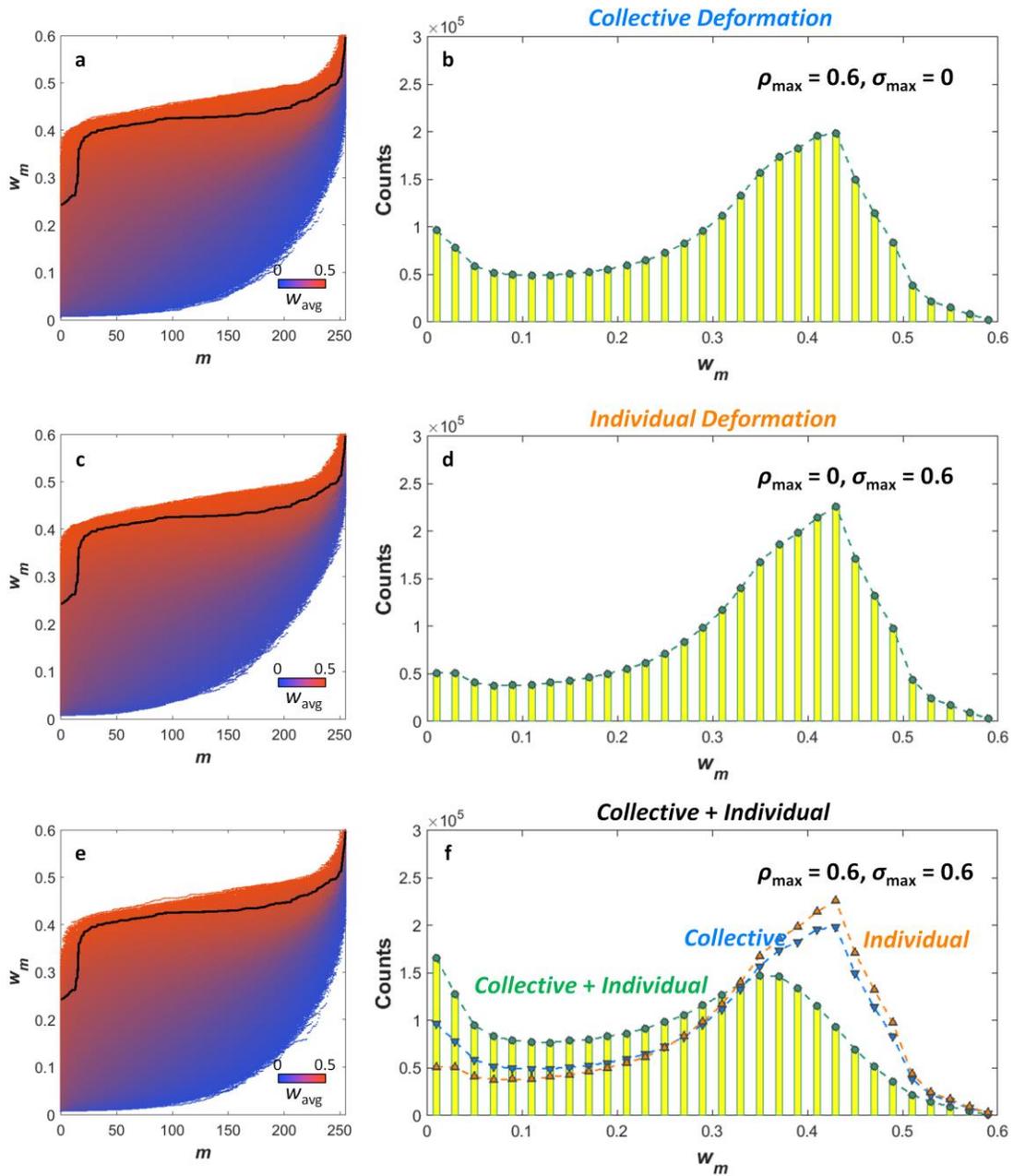

**Fig. S1. Localizations with collective and individual deformations. a, b,** Collective only ($\rho_{max}$ = 0.6, $\sigma_{max}$ = 0), **c, d,** individual only ($\rho_{max}$ = 0, $\sigma_{max}$ = 0.6), and **e, f,** simultaneously collective and individual deformations ($\rho_{max}$ = 0.6, $\sigma_{max}$ = 0.6). **a, c, e,** Localization values. Black solid lines denote the localization data for each realization example. **b, d, f,** Statistical distributions of the localization values $w_m$. Blue and orange dashed lines with symbols in **f** denote the results in **b, d** for comparison. For all cases, $1 \times 10^4$ realizations are considered.

**Note S2. Training process of the D2L and L2D CNNs**

Figure S2 shows the training process of the D2L and L2D CNNs. The accuracy is calculated with the validation dataset after the training of each epoch, whereas the cost function is obtained with the training dataset during the optimization process. The dropout method [2] operates better than the L2 regularization [3] in the training of the D2L CNN, and the L2 regularization provides excellent training performance for the L2D CNN, which actually entails training the L2D2L CNN with the fixed D2L CNN part.

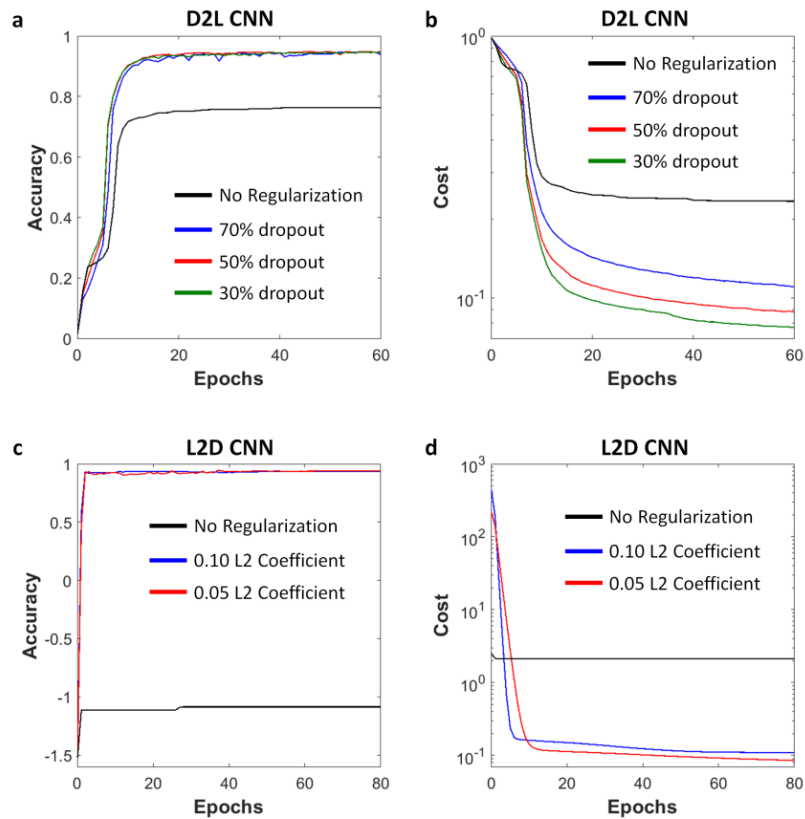

**Fig. S2. Training process of CNNs. a, b,** D2L and **c, d,** L2D CNNs for **a, c,** validation accuracy and **b, d,** training cost functions: **a,** $1 - L_{\text{D2L}}$ for the validation dataset, **b,** $L_{\text{D2L}}$ for the training dataset, **c,** $1 - L_{\text{L2D2L}}$ for the validation dataset, and **d,** $L_{\text{L2D2L}}$ for the training dataset. Each coloured line represents different regularization parameters for the **a, b,** dropout method and **c, d,** L2 regularization. The dropout probability in **a, b** represents the ratio of randomly assigned inactive neurons in the FC layer. A larger L2 regularization coefficient in **c, d** derives stronger regularization for the cost function in TensorFlow [4].

**Note S3. Dependence of power-law distributions on data size**

To examine the reliability of the power-law fitting result in the main text, we analyse the data-size dependence of the power-law exponent $\alpha$ for $(\Delta r)^{-\alpha}$ and the lower bound of the heavy tail $\Delta r_{\min}$. The data size is determined by the number of realizations, where each realization denotes an ML-generated disordered structure composed of 256 atoms. Figures S3a and S3b represent $\alpha$ and $\Delta r_{\min}$ as functions of the realization number, respectively. The result shows that the power-law exponent $\alpha$ is saturated with $\sim 10^1$ (2560 atoms) realizations, while the lower bound of the heavy tail $\Delta r_{\min}$ is saturated with $\sim 10^2$ (25600 atoms) realizations. Notably, even a single realization also has a similar value of $\alpha$ and $\Delta r_{\min}$.

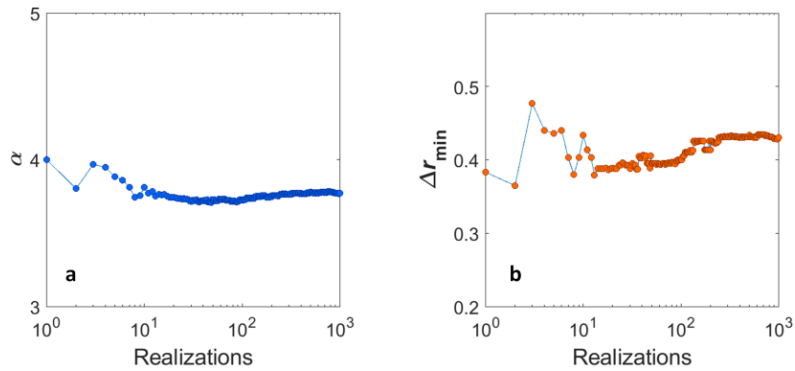

**Fig. S3. Dependence of power-law parameters on the number of realizations. a,** Power-law exponent $\alpha$. **b,** The lower bound of the heavy tail $\Delta r_{\min}$.

**Note S4. Scale invariance in ML-generated materials with different degrees of localization**

In Fig. 4 in the main text, we analyse the microstructural statistics of ML-generated disordered structures, which have an average mode area $w_{avg}$ in the range of $0.20 \leq w_{avg} \leq 0.30$. In this Note, we compare the microstructural statistics of ML-generated structures in different degrees of localization to examine whether the scale invariance is universally observed. Figure S4 shows the microstructural statistics of the seed and ML-generated structures for different ranges of $w_{avg}$. The range of statistical distributions decreases for weaker localization (or larger $w_{avg}$), showing the convergence to unperturbed crystals with a maximum $w_{avg}$. However, a "heavy-tail" distribution of ML-generated structures is always maintained regardless of the degrees of localization, especially when compared to a normal distribution of seed structures.

In Fig. S5, we also apply the analysis based on the maximum-likelihood fitting method with goodness-of-fit tests [5,6] to each case of statistical distributions. Regardless of the degrees of localization, the ML-generated structures present a very similar statistical distribution near its "heavy tail", providing the power-law distribution $(\Delta r)^{-\alpha}$ where $3.24 \leq \alpha \leq 3.92$. Therefore, the scale invariance of ML-generated disordered structures is universally observed for varying degrees of localization, although the position of heavy tails changes to meet the target localization condition.

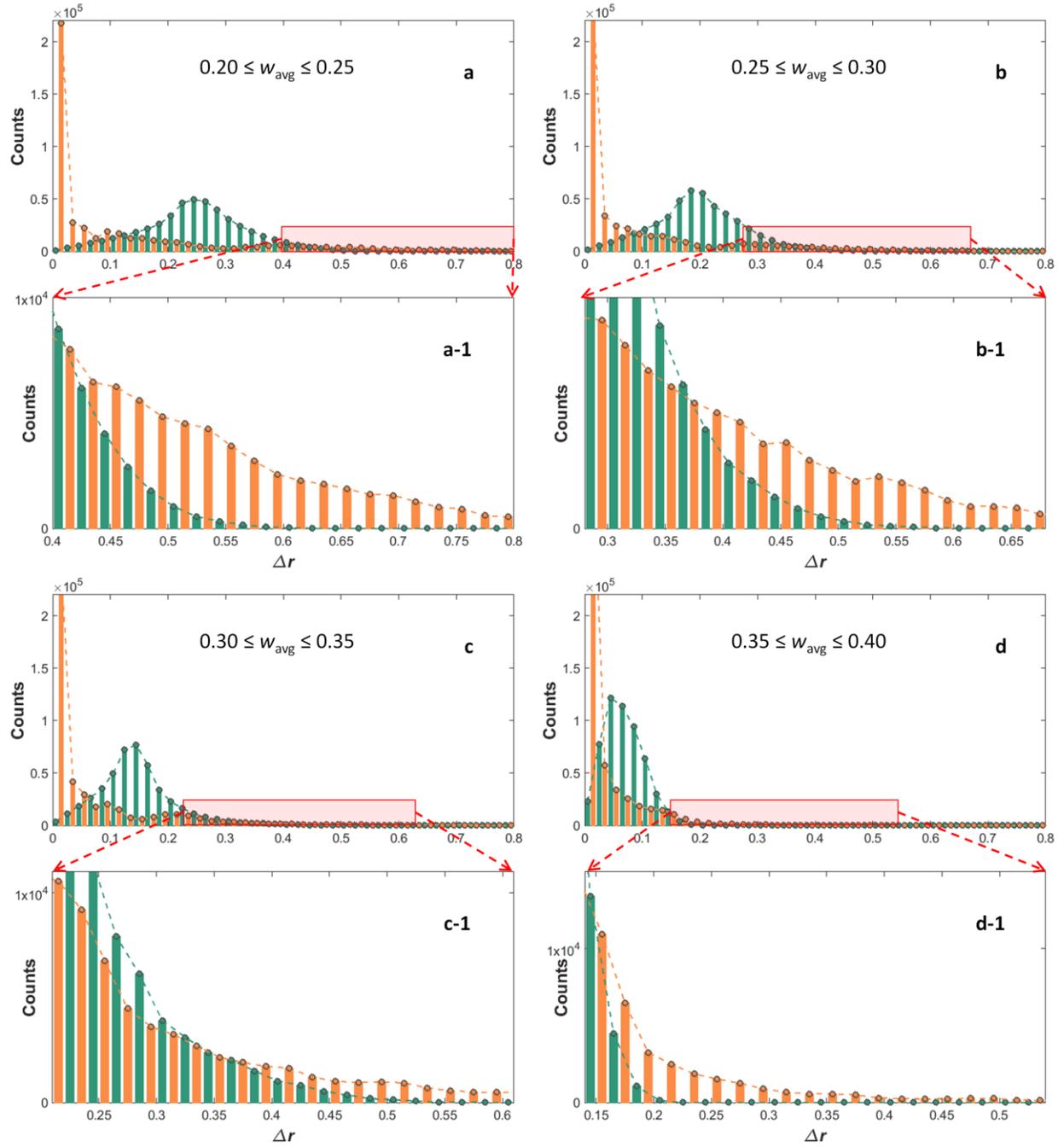

**Fig. S4. Statistical distributions for different degrees of localization.** The strength of the lattice deformation $\Delta r$ is shown for the seed (green) and ML-generated (orange) structures. **a,** $0.20 \leq w_{\mathrm{avg}} \leq 0.25$ (1823 realizations), **b,** $0.25 \leq w_{\mathrm{avg}} \leq 0.30$ (1377 realizations), **c,** $0.30 \leq w_{\mathrm{avg}} \leq 0.35$ (1588 realizations), and **d,** $0.35 \leq w_{\mathrm{avg}} \leq 0.40$ (2123 realizations) in the ML design. Each inset (a-1 to d-1) shows the extended plot near a heavy tail.

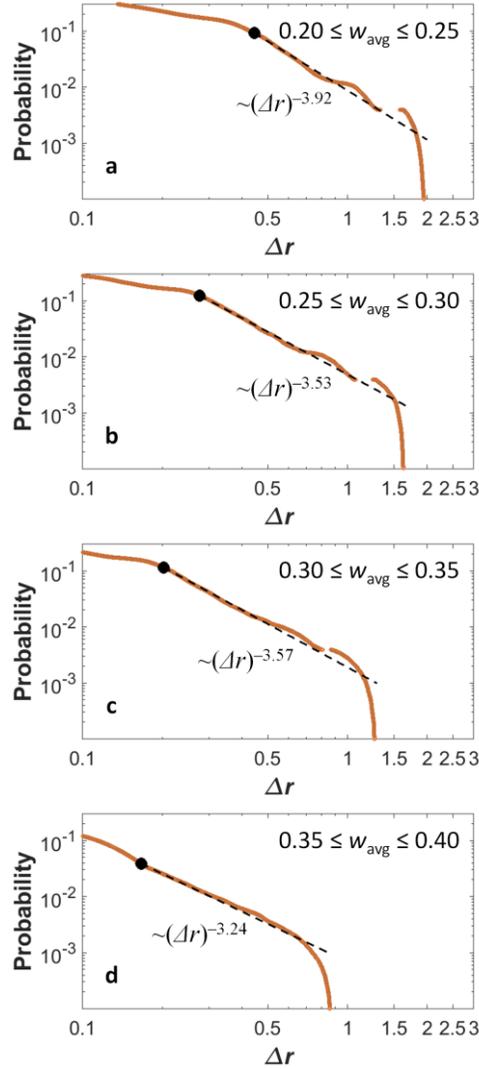

**Fig. S5. Power-law analysis near heavy tails.** Each plot shows the log-log plot of each case in Fig. S4, showing the power-law distribution. The orange line (composed of discretized points) represents the complementary cumulative distribution function (CDF) obtained from the data set in Fig. S4. The black dashed line represents the best power-law fit to the data using the method in [5,6]. The black dot represents the lower bound $\Delta r_{min}$ to the power-law behaviour. **a,** $0.20 \leq w_{avg} \leq 0.25$ (1823 realizations), **b,** $0.25 \leq w_{avg} \leq 0.30$ (1377 realizations), **c,** $0.30 \leq w_{avg} \leq 0.35$ (1588 realizations), and **d,** $0.35 \leq w_{avg} \leq 0.40$ (2123 realizations) in the ML design.

**Note S5. Energy spectra in normal-random and scale-free disorder**

Figure S6 shows the energy spectra with respect to localization values in normal-random (Fig. S6a) and scale-free (Fig. S6b) disordered structures. The set of structures is composed of 5057 different realizations, which achieve the test accuracy over 80% between the target localization from normal-random disordered structures and the true values from ML-generated scale-free disordered structures. Despite very similar localization values (Figs. 3c and 3e in the main text), normal-random and scale-free disordered structures exhibit apparently different energy spectra, wherein there are more spectrally discrete states in scale-free disordered structures. Thus, the proposed ML inverse design method provides the realization of material structures for the target property of "selected" wave quantities (localization). Depending on the setting of input and output data of the CNNs, the result of wave localization can be extended into the independent and systematic handling of other wave quantities, such as bandgap materials with different localization properties [7] and broadband angular scattering with designed spectral responses [8].

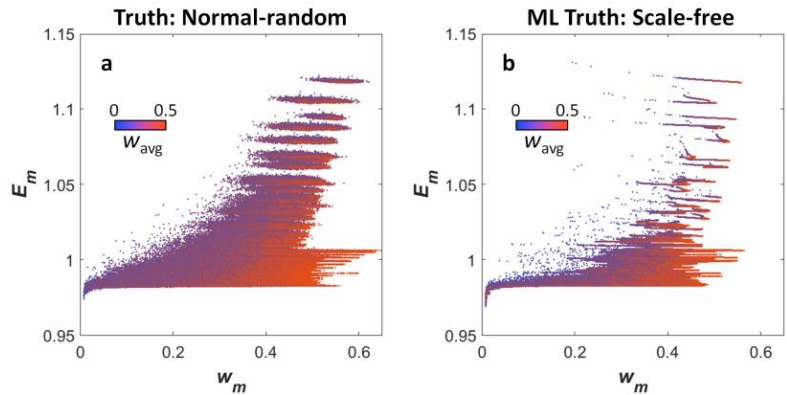

**Fig. S6. Energy spectra of disordered structures with different microstructural statistics.** Localization-energy relations of **a,** normal-random disordered structures obtained from Eq. (5) in Methods and **b,** scale-free disordered structures obtained from the L2D CNN.

**Note S6. Scale invariance in a different ML architecture**

To examine the generality of scale invariance observed in the main text, we investigate another ML architecture: a single pooling layer design. The new architecture, having a comparable number of parameters with that of the original one (~$1.8 \times 10^7$ parameters each in the D2L and L2D CNNs), is shown in Fig. S7. For both D2L and L2D CNNs, the numbers of filters of the convolution layers are set to 16, 32, 64, and 128 in the first, second, third, and fourth layers, respectively. A max pooling layer is used after the fourth convolution layer. The reshaped 1D array is then connected to the FC layer, which has 2048 neurons. We apply the dropout method [2] in the D2L CNN (keeping 60% of FC neurons) and apply the L2 regularization [3] in the L2D CNN (scale parameter: 0.02). All the other conditions are the same as those of the original 3-pooling-layer design. The D2L and L2D2L CNNs achieve the test accuracies ~94.89% and ~95.06%, respectively. We compare the target localizations to the Hamiltonian-calculated true values of ML-generated disordered structures, achieving the good agreement (~84.32%).

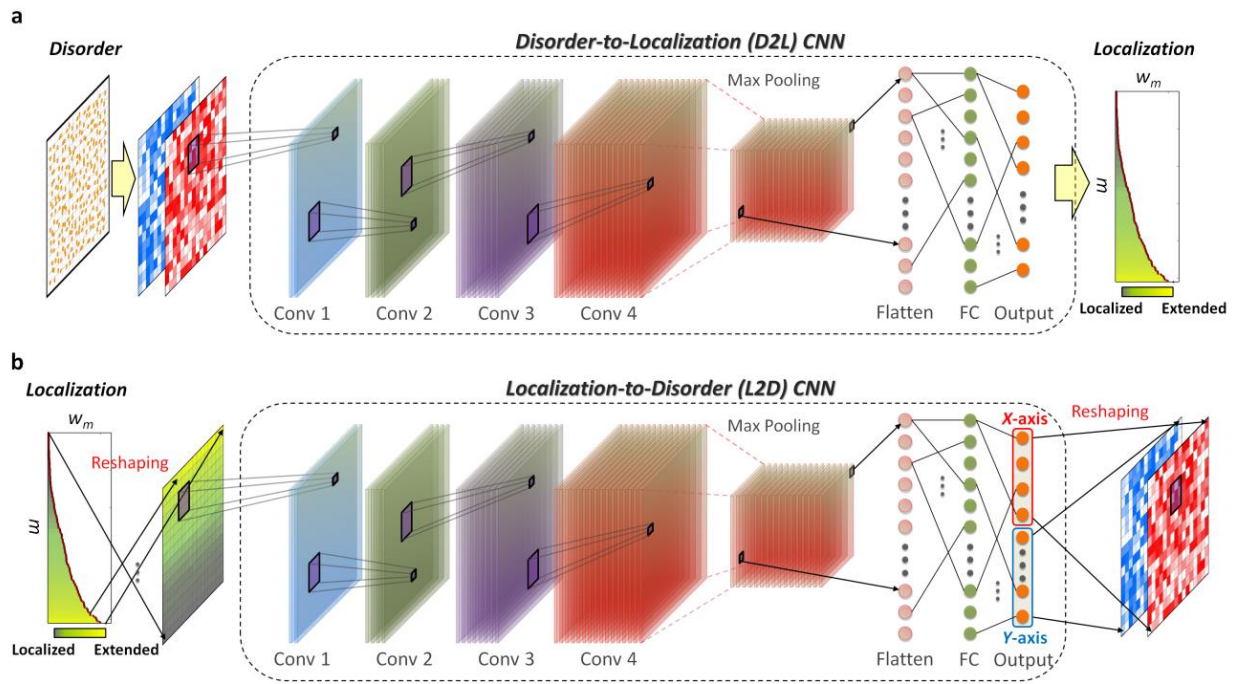

**Fig. S7. ML network structures with a single pooling layer. a,** D2L CNN. **b,** L2D CNN.

Figure S8 compares the microstructural statistics of the seed and ML-generated structures in the original design (Fig. S8a, which is the same as Fig. 4g in the main text) and the new design (Fig. S8b). In both cases, the ML-generated class follows power-law statistics (inset (a-1) and (b-1) each of Fig. S8a,b) and possesses a "heavy-tail" distribution (inset (a-2) and (b-2) each of Fig. S8a,b). However, the shape and range of the "tail" are dependent on the ML architecture with different power-law fitting results: $\sim(\Delta r)^{-3.79}$ in the 3-pooling-layer ML and $\sim(\Delta r)^{-4.09}$ in the 1-pooling-layer ML.

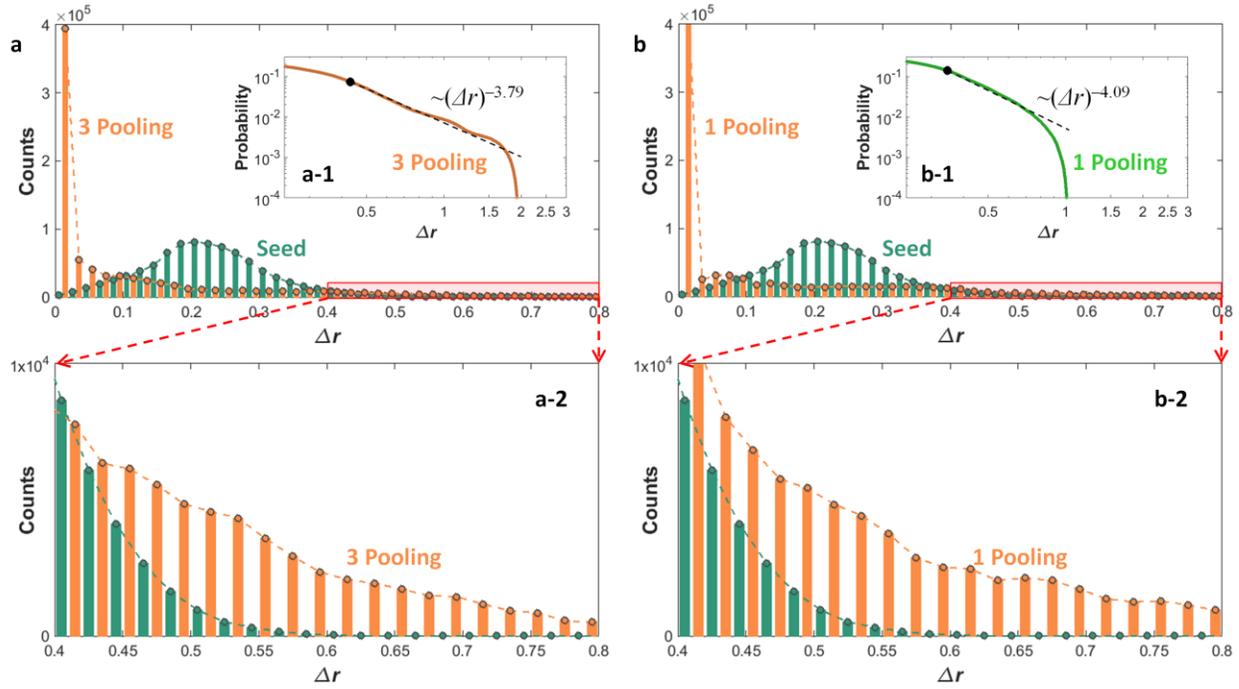

**Fig. S8. Heavy tails in different ML architectures.** Statistical distributions of the strength of the lattice deformation $\Delta r$ in the seed (green) and ML-generated (orange) structures for 3200 realizations satisfying $0.20 \leq w_{\mathrm{avg}} \leq 0.30$ in the 3-pooling-layer ML design: **a,** the distribution of the 3-pooling-layer design, which is Fig. 4g in the main text and is shown for comparison. **b,** the distribution of the 1-pooling-layer design. The first insets **a-1** and **b-1** show the log-log plot. The orange and green lines (composed of discretized points) represents the CDF obtained from the data sets in **a** and **b**. The black dashed lines represent the best fit to the data using the method in [5,6]. The black dot represents the lower bound to the power-law behaviour. The second insets **a-2** and **b-2** show the extended plot near the heavy tails.

Figure S9 represents the comparison of the relationships between ML architectures and ML-generated disordered structures in the 3-pooling-layer (Fig. S9a,b) and 1-pooling-layer designs (Fig. S9c,d). The quantity $W_j = \sum_i [(w_{ji}{}^x)^2 + (w_{ji}{}^y)^2]$ for the weight strengths from the FC layer (2048 neurons) to the output layer (512 neurons) is used to estimate the critical weights in the ML architecture. We note that while the heavy-tailed distribution is always maintained, the decrease of the "tail length" in the 1-pooling-layer design (Fig. S9c versus Fig. S9a) originates from the decreased range of the weight strength $W_j$ (Fig. S9d versus Fig. S9b), which shows that the control of the $W_j$ distribution (or ML architecture) leads to the corresponding alteration of ML-generated structures.

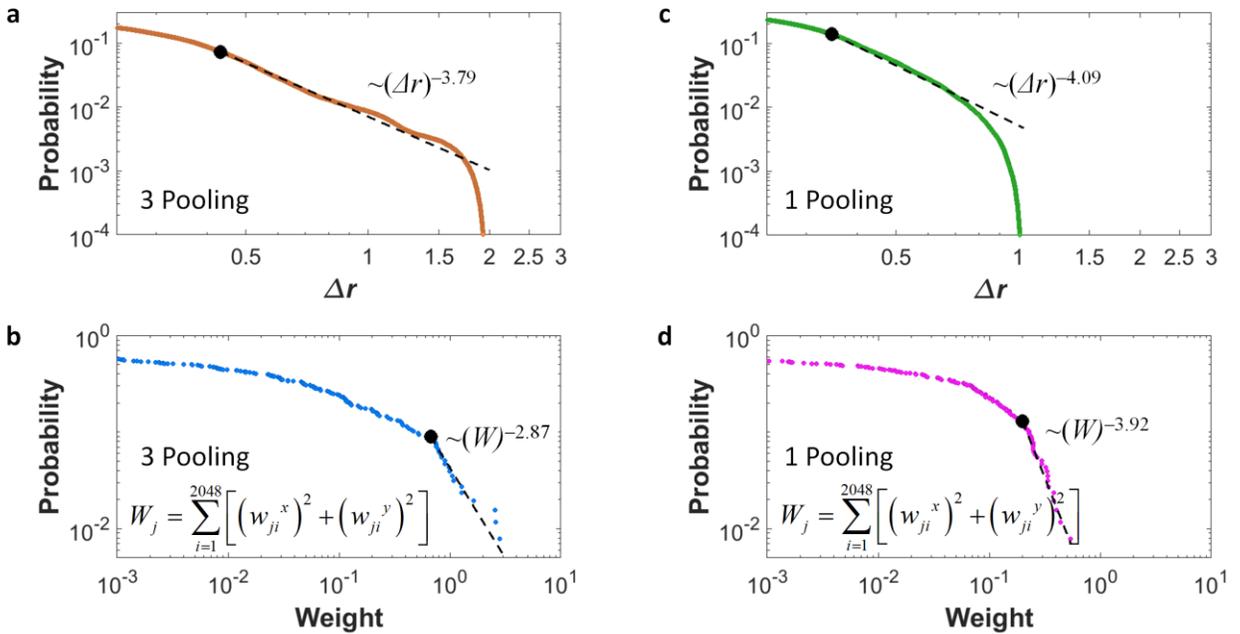

**Fig. S9. Relationships between disordered materials and ML networks for different ML architectures**. **a, b,** 3-pooling-layer ML design and **c, d,** 1-pooling-layer ML design. **a, c,** Power-law fitting of the statistical distribution of $\Delta r$ in ML-generated disordered structures. **b, d,** Power-law fitting of the statistical distribution of the weight parameter $W_j$. **a** and **b** are Figs. 5a and 5b in the main text, respectively, and are shown for comparison.

**References for Supplementary Information**

[1] S. Torquato, *Random heterogeneous materials: microstructure and macroscopic properties* (Springer Science & Business Media, New York, 2002), Vol. 16.

[2] N. Srivastava, G. Hinton, A. Krizhevsky, I. Sutskever, and R. Salakhutdinov, Dropout: a simple way to prevent neural networks from overfitting, J. Mach. Learn. Res. **15**, 1929 (2014).

[3] I. Goodfellow, Y. Bengio, and A. Courville, *Deep learning* (MIT press, 2016).

[4] M. Abadi, A. Agarwal, P. Barham, E. Brevdo, Z. Chen, C. Citro, G. S. Corrado, A. Davis, J. Dean, and M. Devin, Tensorflow: Large-scale machine learning on heterogeneous distributed systems, arXiv preprint arXiv:1603.04467 (2016).

[5] A. Clauset, C. R. Shalizi, and M. E. Newman, Power-law distributions in empirical data, SIAM review **51**, 661 (2009).

[6] J. Alstott and D. P. Bullmore, powerlaw: a Python package for analysis of heavy-tailed distributions, PloS one **9** (2014).

[7] S. Yu, X. Piao, J. Hong, and N. Park, Bloch-like waves in random-walk potentials based on supersymmetry, Nat. Commun. **6**, 8269 (2015).

[8] K. Chung, S. Yu, C. J. Heo, J. W. Shim, S. M. Yang, M. G. Han, H. S. Lee, Y. Jin, S. Y. Lee, and N. Park, Flexible, Angle-Independent, Structural Color Reflectors Inspired by Morpho Butterfly Wings, Adv. Mater. **24**, 2375 (2012).